\documentclass[a4paper,fleqn,usenatbib]{mnras}
\usepackage{multirow}
\usepackage{newtxtext,newtxmath}

\usepackage{color}

\usepackage[T1]{fontenc}
\usepackage{ae,aecompl}
\usepackage{aas_macros}

\usepackage{graphicx,graphics}	
\usepackage{amsmath}	
\usepackage{amssymb}	
\usepackage{natbib}

\title[Azimuthal variations of oxygen abundance profiles ]{Azimuthal variations of oxygen abundance profiles in star-forming region 
 of disc galaxies in the EAGLE simulations}
\author[Solar et al. ]{ Mart\'in Solar$^{1}$, Patricia B. Tissera$^{2}$\thanks{E-mail:
patricia.tissera@unab.cl}, Jose A. Hernandez-Jimenez$^{2}$
\\
$^{1}$Instituto de  F\'isica y Astronom\'ia, Facultad de Ciencias,  Universidad de Valpara\'iso, Gran Breta\~na 1111, Valpara\'iso, Chile.\\
$^{2}$Departamento de Ciencias F\'isicas, Universidad Andr\'es Bello,
700 Fern\'andez Concha, Las Condes, Santiago, Chile. \\
}

\date{Accepted XXX. Received YYY; in original form ZZZ}

\pubyear{2019}

\begin{document}
\label{firstpage}
\pagerange{\pageref{firstpage}--\pageref{lastpage}}
\maketitle


\begin{abstract}

The exploration of the spatial distribution of chemical abundances in star-forming regions in galactic discs provides clues to understand the complex interplay of physical processes that regulate the star formation activity and the chemical enrichment  across a galaxy.
We study the azimuthal variations of the normalized oxygen abundance profiles in the highest numerical resolution run of the Evolution and Assembly of GaLaxies and their Environments (EAGLE) Project at $z=0$. 
We use young stellar populations to trace the abundances of star-forming regions. Oxygen profiles are estimated along different line of sights from a centrally located observer.
 The mean azimuthal variation  in the EAGLE discs are $\sim 0.12 \pm 0.03$~dex~$R_{\rm eff}^{-1}$ for slopes and $\sim 0.12 \pm 0.03$~dex for the zero points, in  agreement with previous works. Metallicity gradients measured along random directions correlate with those determine by averaging over the whole discs although with a large dispersion.
We find a slight trend for higher azimuthal variations in the disc components of low star-forming and bulge-dominated galaxies.
We also investigate the metallicity profiles of stellar populations with higher and lower levels of enrichment than the average metallicity profiles, and we find that  high star-forming region with high metallicity tend to  have slightly  shallower metallicity slopes compared with the overall metallicity gradient.
The simulated azimuthal variations in the EAGLE discs are in global agreement with observations, although the large variety of metallicity gradients would encourage further exploration of the metal mixing in numerical simulations. 

\end{abstract}


\begin{keywords} galaxies: evolution, galaxies: abundances, galaxies: ISM
\end{keywords}


\section{Introduction}

The standard model for disc formation proposed by \citet{Fall1980_1}
is based on the hypothesis of specific angular momentum conservation of the gas as it cools. Numerical simulations have shown that if the condition of angular momentum conservation is
globally fulfilled, discs that satisfy the observed scale relations are formed \citep[e.g.][]{Pedrosa2015_1, Lagos2016_1}. In this context, discs form inside-out and, hence, the star formation starts in the central
region and moves to the outskirts, contributing to set negative metallicity profiles \citep[e.g.][]{Chiappini2001_1, Pilkington2012_1, Tissera2016_2}. 
 The characteristics of the metallicity profiles provide insight into a variety of physical processes related to galaxy evolution such as nucleosynthesis, stellar winds, star formation (SF) and outflows, among others.
These processes can change the oxygen abundance distribution of the interstellar medium (ISM) in a complex way as each of them operates at different time-and-space-scales.
Additionally, secular evolution and radial migration, mergers and galaxy-galaxy interactions are dynamical processes that can also  modify distribution of  chemical abundances in the ISM and the stellar populations (SPs), reshaping the metallicity profiles  \citep{Rupke2010_1, Amorin2012_1, DiMatteo2013_1, Grand2016_1, Molla2016_1, Tissera2016_2,  Sillero2017_1, Ma2017_1, Tissera2019_1}.

In the Local Universe, HII regions in spiral galaxies are known to have negative abundance gradients, on average, \citep[e.g.][]{Searle1971_1, Martin1992_1, Zaritsky1994_1, Kennicutt2003_1, RosalesOrtega2011_1} so that the inner regions are more metal-rich than the outskirts.
However, departures from a single metallicity gradients \citep{SanchezMenguiano2017_1} and the existence of inverted metallicity gradients associated to interacting galaxies have been also reported  \citep[e.g.][]{Rupke2010_1, rosa2014}.
 First observations of nearby spiral galaxies showed no clear indiction of important  azimuthal variations in the metallicity distributions \citep[e.g.][]{Martin1996_1, Cedres2002_1, Cedres2012_1, Kennicutt1996_1, Li2013_1, Zinchenko2016_1}.
Only recently, with the help of Integral Field Spectroscopic (IFS) surveys, it has been possible to analyse the metallicity distributions across the discs in more detail and with a larger statistical significance.
\citet{SanchezMenguiano2017_1} analysed 63 face-on spiral galaxies selected from CALIFA data, finding  differences in the chemical abundances of the arms and interarms regions, with modal values of $-0.013$~dex$R_{\rm eff}^{-1}$  and  $-0.015~$dex$R_{\rm eff}^{-1}$ for flocculent and grand design spirals, respectively. 
Detailed observations of individual galaxies also provide evidences of azimuthal variations \citep{Li2013_1, SanchezMenguiano2016_1, Vogt2017_1, Ho2017_1}.
For example, \citet{Ho2018_1} analysed the oxygen abundance of HII regions in NGC 2997, reporting $\sim 0.06$ dex azimuthal variations in the oxygen abundance, with higher enrichment  in the arm regions. 
Although the reported azimuthal variations are small, they provide information on the regulation of the SF activity and metal mixing process across the discs.

From a theoretical point of view, \citet{Grand2016_1} used high resolution hydrodynamical simulations to follow the evolution of radial flows associated to the spiral arms. These authors  showed how they could produce an 
overdensity of high-metallicity stars in the trailing side of the arms and an overdensity of metal-poor stars on the leading side. 
\citet{DiMatteo2013_1} reported that radial migration induced by a bar can produce azimuthal variation of the old
stellar populations using pure N-body simulations.
\citet{Khoperskov2018_1} analysed the formation of azimuthal metallicity variations using high resolution N-body simulations of discs with no initial metallicity gradients. They found that the different responses to a spiral perturbation of the kinematically
hot and cold stellar populations produce variations in the metallicity distributions. Using analytical 2D chemical models, \citet{Spitoni2018_1} study and quantify the effects of spiral arm density fluctuations
on the azimuthal metallicity variations. These authors report an azimuthal variation  of the metallicity gradients of the order of $\sim 0.1$ dex.

Numerical simulations that include chemical models, such as the simulations from the Evolution and Assembly of GaLaxies and their Environment  (EAGLE) Project \citep{Crain2015_1, Schaye2015_1}, are powerful tools to contrain the subgrid models for the evolution of baryons. Chemical patterns
give additional information on the processes that
regulate the star formation activity and redistribute the angular momentum and mass in galaxies as well as on the impact of inflows and outflows. Because they  are usually not
used to fix the free parameters of the algorithms of the subgrid physics, their analysis provide   constraints to them. The chemical elements are often distributed within the nearby regions of the stellar sources \citep[][]{Mosconi2001_1}. Unless other mechanisms are included, there  will no further exchange of material within the minimum resolved volumes. Metal diffusion is expected to contribute to mitigate this issue at the expense of introducing an extra free parameter, the metal diffusion coefficient \citep{Greif2009,Pilkington2012_2}. The characteristics of the metal distribution and the gradients will be affected by the efficiency of the mixing and/or diffusion of chemical elements. This is an important issue considering that the cooling rates depend strongly on the chemical  abundances of the ISM for example. The more detailed observational data that are being gathered by IFS surveys open the possibility to perform more detail comparisons with models and hence, we expect the analysis presented in this paper to
serve as a benchmark for future improvements. 

In this work, we analyse the azimuthal variation of oxygen abundances of the discs identified from a set of simulated galaxies extracted from the high-resolution (25 Mpc)$^3$  volume of  the EAGLE Project. The EAGLE simulations have proven to reproduce global properties of galaxies such as the mass-metallicity relation \citep{derossi2018}, the fundamental relations of early type galaxies \citep{Lagos2018_1,Trayford2019_1,Rosito2019}, and  the locally-resolved scale relation between star formation and metallicity  \citep{Trayford2019_2}, among others. 
In particular, these simulations provide a large sample of galaxies with different formation histories, which
opens the possibility to assess the existence of azimuthal metallicity variations.
This work is based on the galaxy catalogue built  by \citet{Tissera2019_1} from which disc galaxies were selected for the higher resolution run of 25 Mpc cubic-side volume. In this paper, we analyse the azimuthal variations of the oxygen abundances in star-forming regions in the disc
components at $z=0$.

This paper  is organised as follows. Section 2 describes the simulations and methods implemented to select the analysed stellar populations.
Section 3 presents the analysis of the azimuthal variations of the chemical abundance distributions. In Section 4 we summarises our main results. The Appendix provide information on the impact of the numerical dispersion in the metallicity distribution.
 

\section{Simulations} \label{sec:2}

The EAGLE Project\footnote{We use the publicly available database by \citet{McAlpine2016_1}.} comprises cosmological hydrodynamical simulations consistent with a $\Lambda$-CDM universe, performed assuming $\Omega_{\rm \Lambda} = 0.693$, $\Omega_{\rm m} = 0.307$, $\Omega_{\rm b} = 0.04825$, $h = 0.6777$ ($H_{\rm 0} = 100$ h km s$^{\rm -1}$ Mpc$^{\rm -1}$), $\sigma_{\rm 8} = 0.8288$, $n_{\rm s} = 0.9611$, and $Y = 0.248$ \citep{Planck2014_1}.
The simulations were performed with an enhanced version of the  {\small GADGET-2} code \citep{Springel2005_1} that includes a modified hydrodynamics solver and time stepping by the ANARCHY model \citep{Schaller2015_1}.

The code tracks the chemical enrichment of eleven chemical elements produced by mass loss by intermediate-mass asymptotic giant branch, Type Ia and II supernovae as well as mass winds by massive stars \citep{Wiersma2009_2}.
A \citet{Chabrier2003_1} initial mass function is adopted.
It also includes radiative cooling and a photo-heating models as described in \citet{Wiersma2009_1}.
The energy feedback model from stellar sources is implemented stochastically \citep{DallaVecchia2012_1} and was calibrated to reproduce the stellar mass function and galaxy sizes at $z=0$.
More details of the simulations and the implemented subgrid physics can be found in \citet{Crain2015_1}.

For this study, we use the re-calibrated simulation (Ref-L025N0752) representing a volume of $25$ Mpc comoving box side, resolved by $752^{\rm 3}$ initial particles.
 The mass resolution is $2.26 \times 10^{\rm 5}$ M$_{\rm \odot}$ and $1.21 \times 10^{\rm 6}$ M$_{\rm \odot}$ for the initial gas and dark matter particles, respectively. 
The  maximum proper gravitational softening is $0.35$pkpc.
 
The EAGLE project also provides smoothed chemical abundances, which are estimated by applying a kernel function used to calculate
smoothed values by using the information of the nearby regions. The smoothed abundances are used to estimate the cooling rates in the EAGLE simulations. Hence, in order
to check their impact and for sake of consistency, we repeated the whole  analysis of the metallicity distributions by using the smoothed abundances in order to assess 
if the results depend on this mixing. We detect  no significant change in the 
azimuthal variations  neither in the trends reported by using the non-smoothed abundances. Selected trends obtained using the smoothed abundances are included in the Appendix A as examples. 

\subsection{Galaxy sample} \label{sec:2.1}

The simulated galaxy sample is selected from the catalogue of \citet{Tissera2019_1}, where a dynamical criteria based on
the angular momentum context and binding energy was applied to identify the disc components \citep{Tissera2012_1}.
The angular momentum content of each particle is quantified by  $\epsilon = J_{\rm z}/J_{\rm z,max} (E)$, where $J_{\rm z}$ is the angular momentum and $J_{\rm z,max} (E)$ is the maximum $J_{\rm z}$ over all particles at a given binding energy $E$.
Only stars with $\epsilon > 0.5$ are taken into count to define the stellar disc component. 
Stars particles that are not rotational supported are considered to belong to the spheroidal component and are not analysed in this work \citep[see][]{Rosito2019}.
 
For the purpose of our analysis, we define two sets of SPs:  young stars which are selected to have ages smaller  than 2 Gyr and  super-young stars, chosen to have ages younger than 0.5 Gyr.
For each simulated galaxies, the star formation rate (SFR),  the specific star formation rate (sSFR), the  half-mass radius ($R_{\rm eff}$), defined as the one that encloses half the  mass of the young (super-young) stellar discs are estimated.
Galaxy morphology is defined by  the ratio between the total stellar mass of the discs and the total stellar mass of a galaxy (D/T). The abundance ratio  12 +\ log \ (O/H) will be used to quantify the level of enrichment of the simulated SPs (hereafter, we will also use the term 'metallicity' to refer to the oxygen abundances).

We follow previous works that showed that young SPs can be taken as tracers of the chemical abundances of the star-forming regions in the ISM \citep{Gibson2013_1, Tissera2016_2}.
We adopt the same criteria considering the two mentioned age thresholds which will be used for different aspects of the analysis. Only those discs with more than 1000 young (super-young) stars will be analysed. This  condition is adopted because the estimation of the metallicity gradients in different azimuthal direction requires to be able to numerically resolve all of them with a
suitable number of star particles. The final samples comprise  106 and 42 discs (sampled by young stars and super-young stars, respectively).
The selected galaxies have stellar masses in the range [$10^9,10^{10.8}$] M$_{\sun}$ and  SFR in range $[0.1,6]$ M$_{\sun} yr^{-1}$.

\begin{figure*}
\resizebox{15cm}{!}{\includegraphics{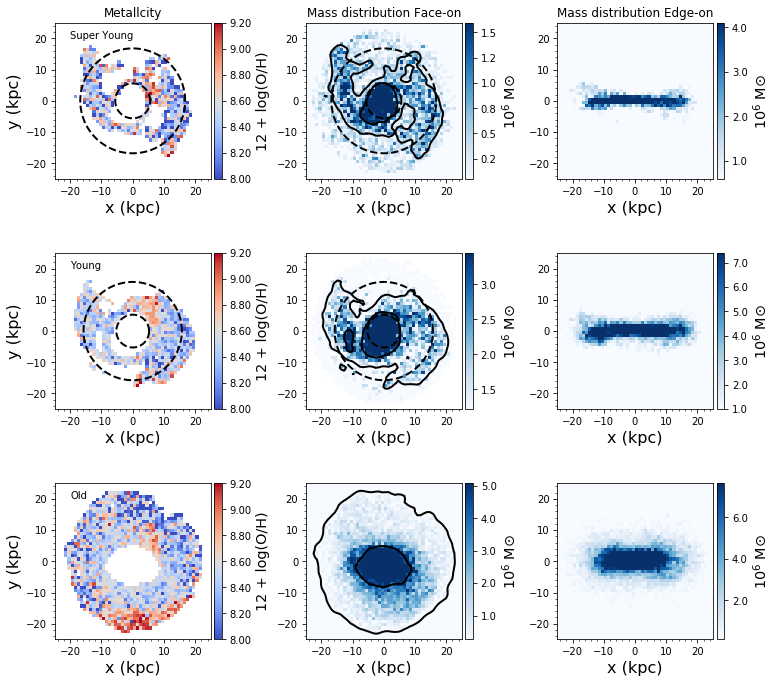}}
\caption{ Metallicity and stellar mass distributions of a galaxy with well-defined spiral arms from our selected sample.
Lelf panels: face-on projections of the oxygen abundance
 (12 \ + \ log \ (O/H)) for the super-young (top), young (middle) and old (bottom) populations. Middle panels: face-on projections
 of stellar mass density of star particles for  super-young (top), young (middle) and old (bottom) populations. The inner and outer black circles represent 0.5 and 1.5 $R_{\rm eff}$, respectively. The black
 contours represent mass iso-densities   chosen to highlight the arm
and disc structures for young  (top and middle) and old (bottom) populations, respectively.
The face-on metallicity projections were masked inside of the mass iso-density contours shown in the middle panels.  Right panels:
 the same as middle panels but for edge-on projections
 of stellar mass density.}
\label{fig:1}
\end{figure*}

In order to illustrate the distribution of both metallicity and mass
of the SPs studied, Fig.~\ref{fig:1} shows the face-on projections of metallicity (left-panels), the face-on (middle-panels) and edge-on (right-panels) projections of the stellar mass density for the  super-young, young and old  (with ages $>2$ Gyr) SPs. To create these maps, we used grids of 50$\times$50 pixel on images with a physical size of 50$\times$50 kpc. For the face-on metallicity projection, each pixel  was assigned the median value along
z-axis, while for the face-on and edge-on stellar mass density projections were assigned the sum of the stellar mass along z- and y-axes, respectively. 
As expected both super young and young SPs are good tracers of the spiral arms as denoted by the iso-density contours on the face-on images and are distributed in thin disc as shown in edge-on projections, while the old SP traces the disc and is distributed in a thick disc (see face-on and edge-on projections, respectively). The metallicity distributions of super-young and young SPs have larger contribution of high metallicity SPs  (12 \ + \ log \ (O/H)$>8.5$, red points), while the old SPs  seem to have more homogeneous values around of 12 \ + \ log \ (O/H)$\sim 8.5$ (white points). The different spatial and numerical distributions of  metallicity  among the SPs will be quantified in more detail in the following sections. 

\section{Azimuthal variations of the oxygen profiles}

In this section, we quantify the azimuthal variations of the oxygen abundance gradients by estimating the metallicity profiles along different directions on the disc plane from an observed located at the galactic centre. Each galaxy is rotated so that the total angular momentum is located along the z-axis. The stellar discs are projected along the z-axis. 
For simplicity, the azimuthal variations are obtained by tessellating in six equal sub-regions the projected  stellar mass distributions, along the main axis of rotation. Each of the resulting regions covers 60$^{\rm o}$ of the total disc (360$^{\rm o}$). The  radial metallicity profiles are obtained by estimating the median oxygen abundance values in  radial intervals, each one enclosing  the same number of star particles\footnote {A total of 20 radial intervals are defined in each disc resolved with N particles, within $[0.5, 1.5]R_{\rm eff}$. The  widths are determined by the distributions of n particles, where n=N/20.}. We only considered sub-regions with at least 200 young SPs, so  each of them had enough star particles to estimate the metallicity profiles. In order to have enough number of SPs in each subregions, in this section, we  work with the young SPs (ages $< 2 $Gyr).

Linear regression fittings to the metallicity profiles are constructed by applying the least trimmed squares (LTS) robust method that considers errors in the y-variable, leaving out possible outliers \citep{Rousseeuw2006_1, Cappellari2013_1}. The errors correspond to three times the bootstrap errors estimated for  each radial interval \footnote{Bootstrap errors are used instead of the standard dispersion in each interval because the analysis of the metallicities distributions shows that the dispersions have a systematic skewness with a  lower metallicity tail  at all radius}.
A linear regression of the form: $y=b(x-x_{\rm 0}) + a$ is used in all cases.
It should be noted that the metallicity profiles are normalized by $R_{\rm eff}$, before applying the LTS fits over the radial range $[0.5, 1.5]R_{\rm eff}$.

After this procedure, six slopes ($\nabla$) and six zero points ($ZP$) for each galaxy are obtained.  We estimated the ZPs at the interception with $x=0$. The median  azimuthal slopes ($\nabla_{\rm M}$) and  zero points ($ZP_{\rm M}$) were estimated  from the linear regressions calculated for the  six sub-regions.
Their respective standard deviations are obtained by applying a bootstrap technique ($\sigma_{\rm \nabla}$ and $\sigma_{\rm ZP}$). Additionally, the overall metallicity gradients were also calculated by using concentric
radial averages  ($\nabla_{\rm T}$ and $ZP_{\rm T}$).  In summary, for each simulated galaxy, we have four parameters: $\nabla_{\rm T}$, $\sigma_{\rm \nabla}$, $ZP_{\rm T}$, and $\sigma_{\rm ZP}$.

\begin{figure*}
\resizebox{17.5cm}{!}{\includegraphics{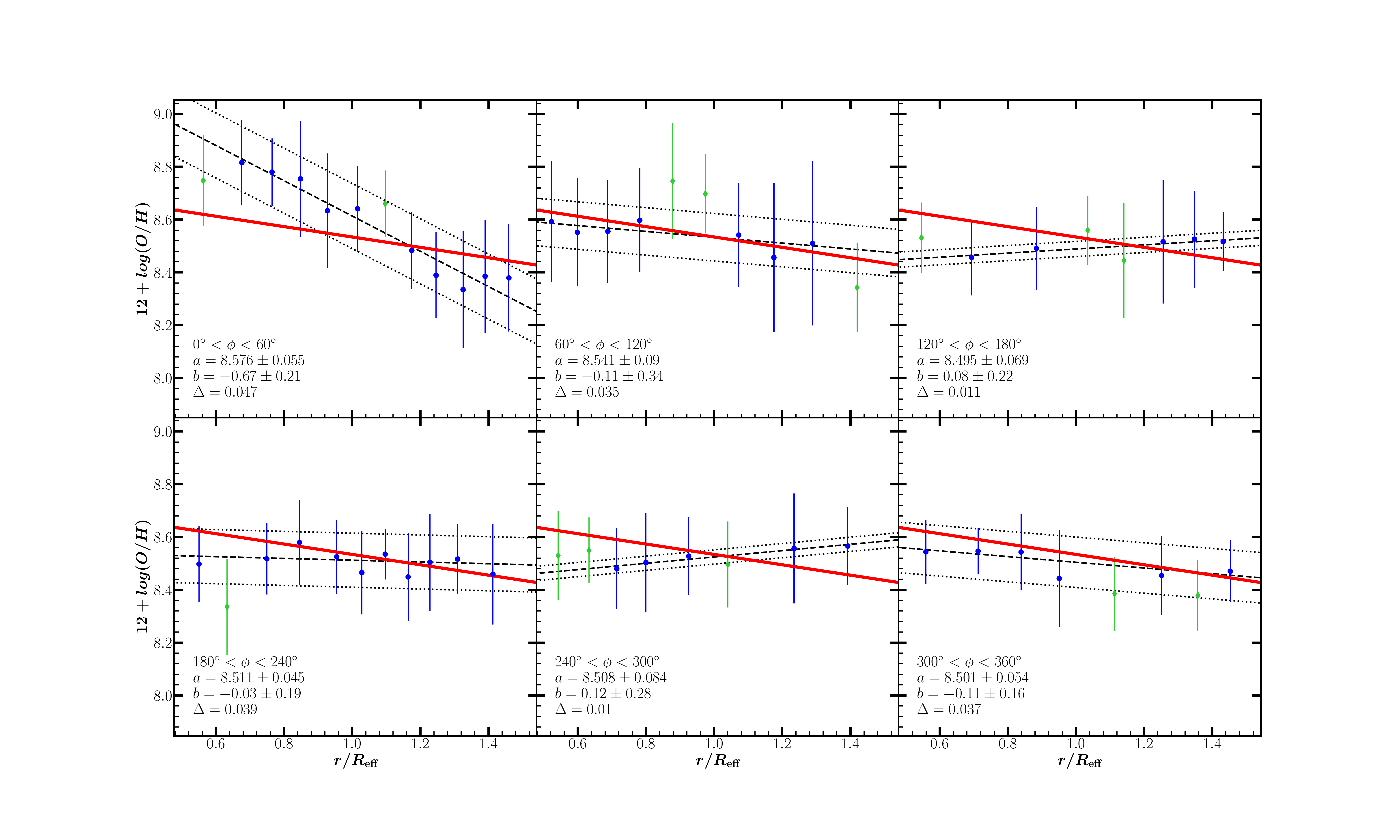}}
\caption{Oxygen abundance profiles of the SF regions located in each of the six sub-regions defined to assess the azimuthal variations for typical galaxy  (black dashed lines). The error bars denote three times the bootstrap errors. For comparison, the
global averaged profile is also shown (red thick line). The standard metallicity deviations in each radial  are included (only blue points are considered for the LTS fist while green symbols are classified as outliers \citep{Rousseeuw2006_1, Cappellari2013_1}). 
The black dotted lines are the linear fits of LTS $\pm$ $ 2.6\triangle$(standard deviation).
}
\label{fig:3}
\end{figure*}

Figure~\ref{fig:3} shows the metallicity profiles obtained for each of the six sub-regions of the a typical simulated disc (black, dashed lines) compared to the 
 overall metallicity profile for the whole disc (red, solid lines) as an example.
As it can appreciate from Fig.~\ref{fig:3},  there are clear azimuthal variations  of the slopes and zero points of the metallicity profiles of each sub-region.

\begin{figure}
\resizebox{9cm}{!}{\includegraphics{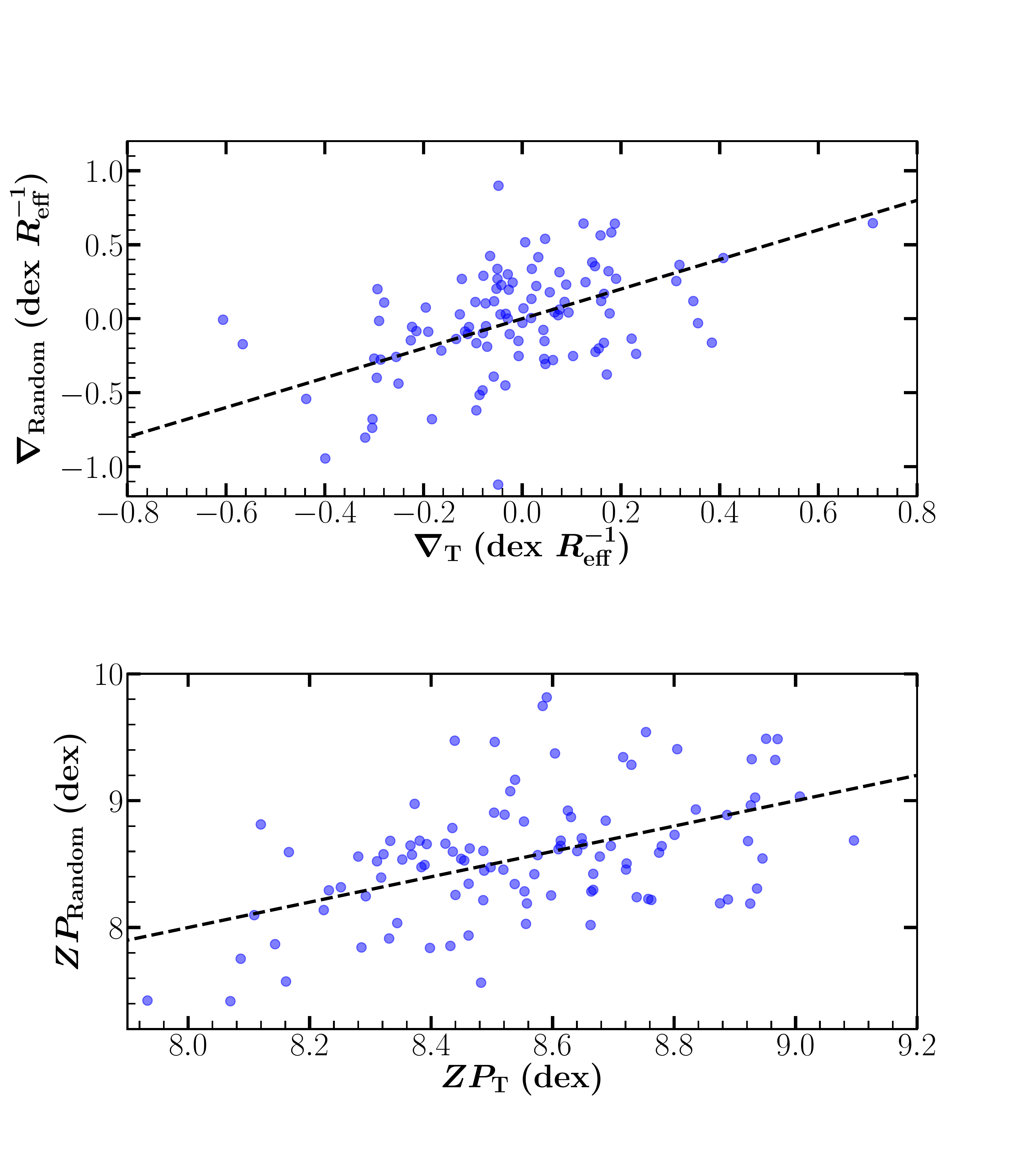}}
\caption{Metallicity gradient $\nabla_{\rm Random}$ estimated along an  azimuthal sub-sample selected at random from the six defined ones as a function of the global metallicity slopes $\nabla_{\rm T}$ (upper panel) and
the corresponding zero points $ZP_{\rm Random}$ as a function of the global ones $ZP_{\rm T}$ (lower panel).}
\label{fig:4}
\end{figure}

 Figure~\ref{fig:4} displays the metallicity gradients and zero points estimated along one of the defined azimuthal subsamples selected at random, $\nabla_{\rm Random}$ and $ZP_{\rm Random}$, as a function of $\nabla_{\rm T}$ and $ZP_{\rm T} $.
The gradient and zero point of the metallicity profiles taken at random from our set of six directions correlates with those obtained
from the radial averages over the whole  discs, indicating that even if there is azimuthal variations, the information store in the overall
profiles can be recovered by using measures along a certain
direction. However, as it can be seen from this figure,  both relations show  large dispersion.  The Spearman coefficient of correlations are  $r \sim 0.46$ and $r \sim 0.43$ for the slopes and zero points,  respectively. This suggests that the azimuthal variations are significant.

\begin{figure}
\resizebox{9cm}{!}{\includegraphics{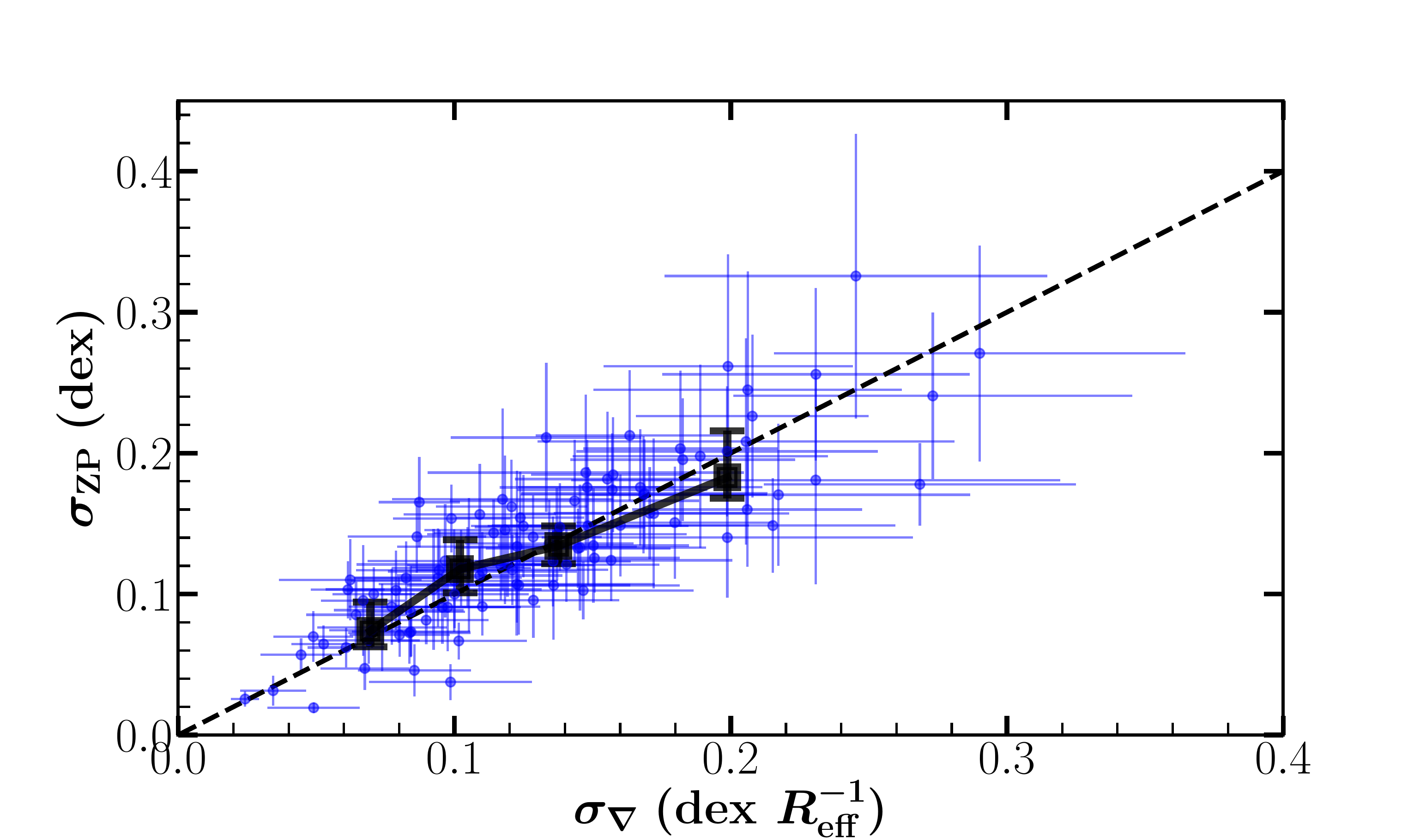}}
\caption{ Median values of the standard deviations of the azimuthal zero points $\sigma_{\rm ZP}$ as a function of those of $\sigma_{\rm \nabla}$ (black squares). The black error bars
represent the 25 and 75 percentile. The blue small dots and blue error bars
represent the data for individual simulated galaxies and the corresponding errors obtained by applying a bootstrap sampling technique over the six azimuthal subregions. }
\label{fig:5}
\end{figure}

The relation between the azimuthal dispersions of the slopes $\sigma_{\rm \nabla}$ and those of  the zero points  $\sigma_{\rm ZP}$ is depicted in the Fig.~\ref{fig:5}.
As can be seen there is a trend for discs with larger $\sigma_{\rm \nabla}$ to also have larger azimuthal variations of the zero point. The error bars are estimated by using a bootstrap technique.  The medians values are  $\sigma_{\rm \nabla} \sim 0.12 \pm 0.03$ ~dex~$R_{\rm eff}^{-1}$  and $\sigma_{\rm ZP} \sim 0.12  \pm 0.03$ dex, respectively.

\begin{figure}
\resizebox{9cm}{!}{\includegraphics{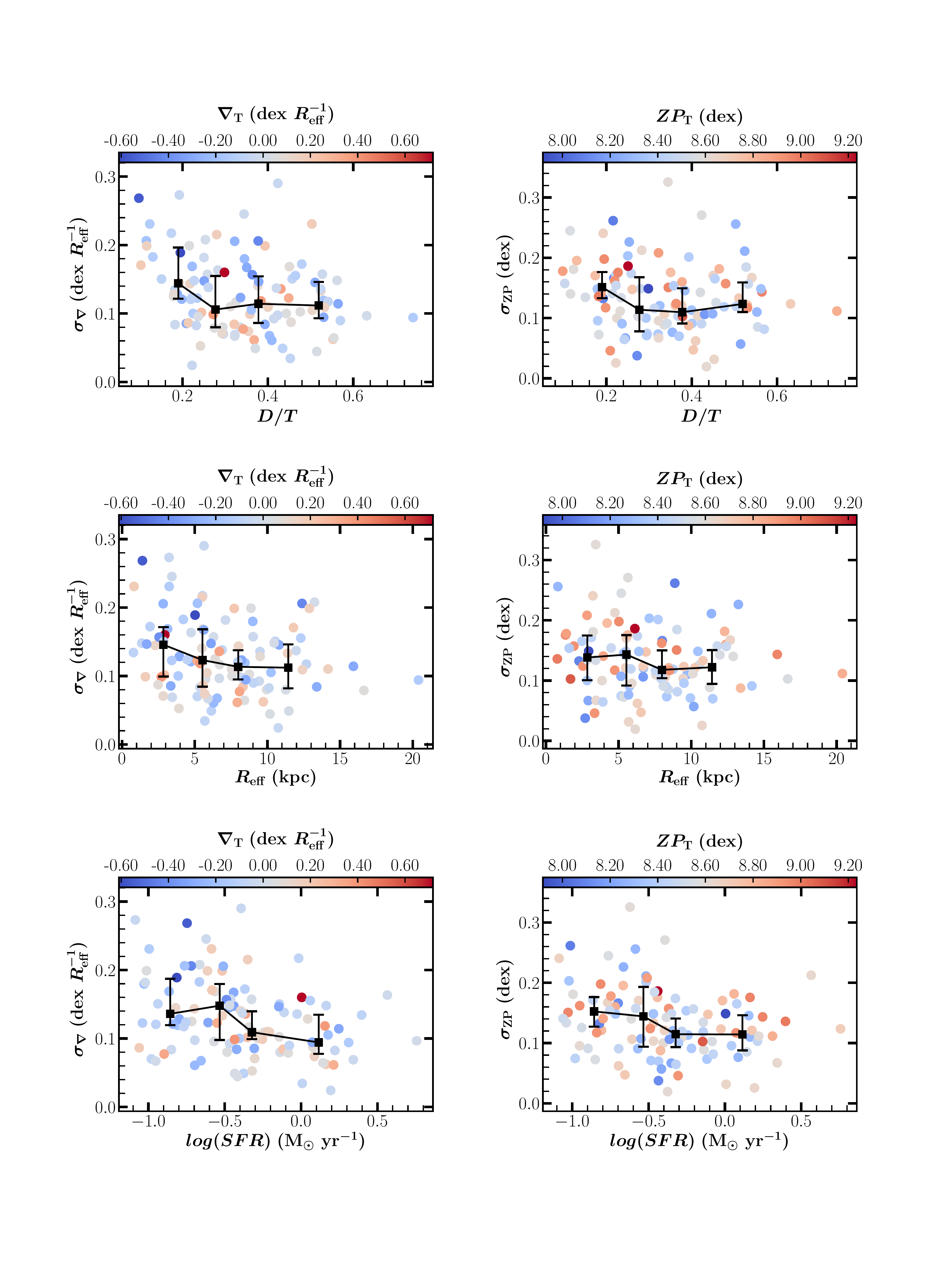}}
\caption{ $\sigma_{\rm \nabla}$ (left panels) and  $\sigma_{\rm ZP}$ (right panels) as a function of $D/T$ (upper  panel), $R_{\rm eff}$ (middle  panel), and SFR (lower  panel). The colour bars show $\nabla_{\rm T}$ (left panels) and $ZP_{\rm T}$ (left panels). Median values are shown (black squares) with  error bars representing the 25 and 75 percentiles.}
\label{fig:6}
\end{figure}

To explore the origin of the azimuthal variation of the metallicity profiles,  Fig.~\ref{fig:6} shows $\sigma_{\rm \nabla}$ and $\sigma_{\rm ZP}$ as a function of $D/T$ (upper panels), $R_{\rm eff}$ (middle panels), and $\rm  log SFR$ (lower panels) coloured by $\nabla_{\rm T}$ (left panels)  and  $ZP_{\rm T}$ (right panels).
The median values and the 25 and 75 percentiles (error bars) are also included.
We find  weak trends of the azimuthal dispersions in the gradients as a function of $D/T$ and SFR. These indicate 
slightly higher azimuthal variations of the metallicity distributions in discs of more bulge-dominated systems and low SFR galaxies.  We would like to point out that  the larger azimuthal variation of the metallicity distributions detected for galaxies with low star- formation activity  could be also affected by  numerical resolution, and hence, these trend should be confirmed by using higher resolution simulations.  No clear trend is found for the 
azimuthal variations of the zero points of the metallicity profiles. 
There are no clear trends of both dispersions with  $R_{\rm eff}$.

The fact that disc-dominated galaxies show less azimuthal dispersion in the metallicity gradients could be associated to their more quiescent evolution history where the ISM got progressively enriched and the stellar distributions were not strongly disturbed.
On the other hand, bulge-dominated galaxies had a larger probability to have experienced mergers or interactions in the  past  as shown by \citet[][]{Tissera2019_1}.
 We speculate that these mechanisms could have disturbed the stellar distributions and hence, increase the azimuthal dispersion, forming a bar for example. The impact would depend on the merger/internaction parameters and 
 how close in time from $z\approx 0$ these event had taken place. A more detailed analysis of possible mixing processes would require better space and temporal resolutions.  Well-defined disc galaxies ($D/T \ge 0.5$) in the EAGLE simulation have azimuthal variations
in the metallicity gradients of $\sim 0.1 $ dex R$_{\rm eff}$.  This value is in agreement with the results reported by  \citep{Spitoni2018_1} for spirals using chemo-dynamical model are analysed.

\section{Oxygen profiles in arm and interarm regions}

\begin{figure}
\resizebox{9cm}{!}{\includegraphics{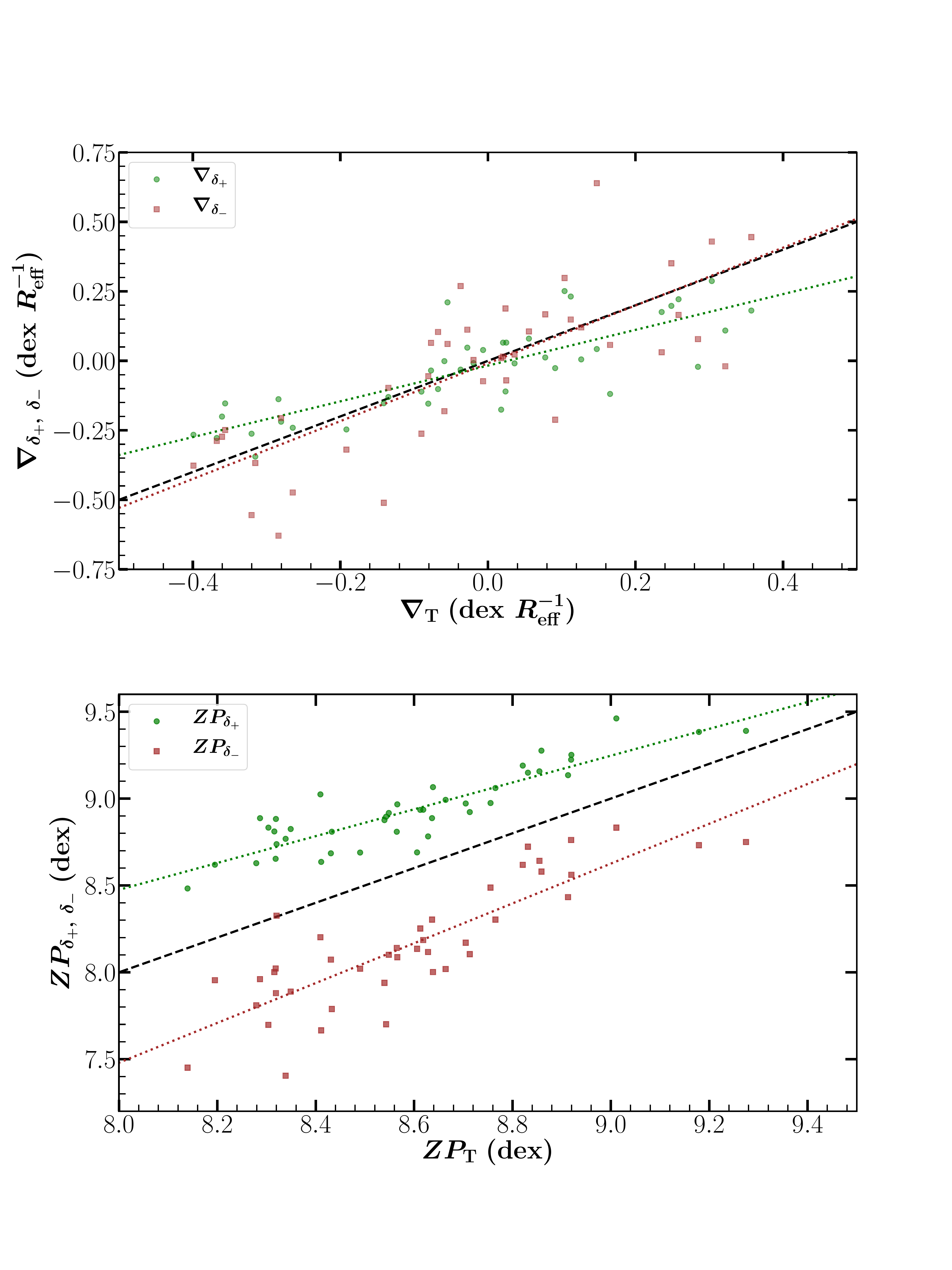}}
 \caption{Upper panel:$\nabla_{\rm \delta_{+}}$ (green dots) and $\nabla_{\rm \delta_{-}}$ (red squares) as a function of $\nabla_{\rm T}$ estimated for the whole distribution of super-young stars. Lower panel:$ZP_{\rm +}$ (green dots) and  $ZP_{\rm -}$ (red squares) as a function of  $ZP_{\rm T}$. The black dashed line corresponds to 1:1 relation in both plots. 
}
\label{exdef}
\end{figure}

Another approach for studying  the azimuthal metallicity variations is the analysis the residual metallicity distributions. 
With the super-young SPs, we estimate the overall metallicity profiles and  obtained $\nabla_{\rm T}$ and $ZP_{\rm T} $,  following the same procedure explained in the previous 
section. Then, the SPs that have an excess or deficit of oxygen abundances with respect to the overall metallicity profile  are identified by defining the 
residues ($\delta$),

\begin{equation} \label{eq:1}
\delta (r_{i}) = Z_{\rm real}(r_{i}) - Z_{\rm fit}(r_{i})
\end{equation}

where $Z_{\rm real}(r_{i})$ is the metallicity of a given super-young SPs located at a galacto-centric distance $r_{i}$ and $Z_{\rm fit}(r_{i})$ is the metallicity it should have at that $r_{i}$, according to the overall linear fit. We note this procedure is applied to all selected stars in order to build up the metallicity residual maps.  We choose to use the fitting relations as  reference values because this is the usual way of quantifying the metallicity distributions at any redshift \citep{carton2018} and they allow us to have a reference value to define the residuals for each individual SP. 
Hence, each super-young SP is classified as having a metallicity excess ($\delta_{+}$) or deficit ($\delta_{-}$)  with respect to the overall  level of enrichment.
The radial metallicity profiles of the super-young SPs with  $\delta_{+}$  and  $\delta_{-}$ are estimated following the same procedure explained in the previous section ($\nabla_{\rm \delta_{+}}$ and  $ZP_{\rm \delta_{+}}$  and  $\nabla_{\rm \delta_{-}}$ and  $ZP_{\rm \delta_{-}}$, respectively). In Fig. ~\ref{exdef}, we show the gradients and zero points  determined by the regions with excess versus those with  deficitwith respect to
the overall metallicity profiles. As can be appreciated there is a slight trend for the regions with excess metallicity to show weaker gradients. From this figure,
it can be appreciated a systematic shift for the zero point values
of the excess and deficitprofiles, with respect to the $ZP_{\rm T}$ of the
global metallicity profiles, respectively. The difference between the
zero points of the regions with excess and deficit metallicities with
respect to the median values is smaller for the galaxies with higher
metallicity. This suggests a higher mixing of chemical elements or/and
a prolonged SF history.

\begin{figure}
\resizebox{9cm}{!}{\includegraphics{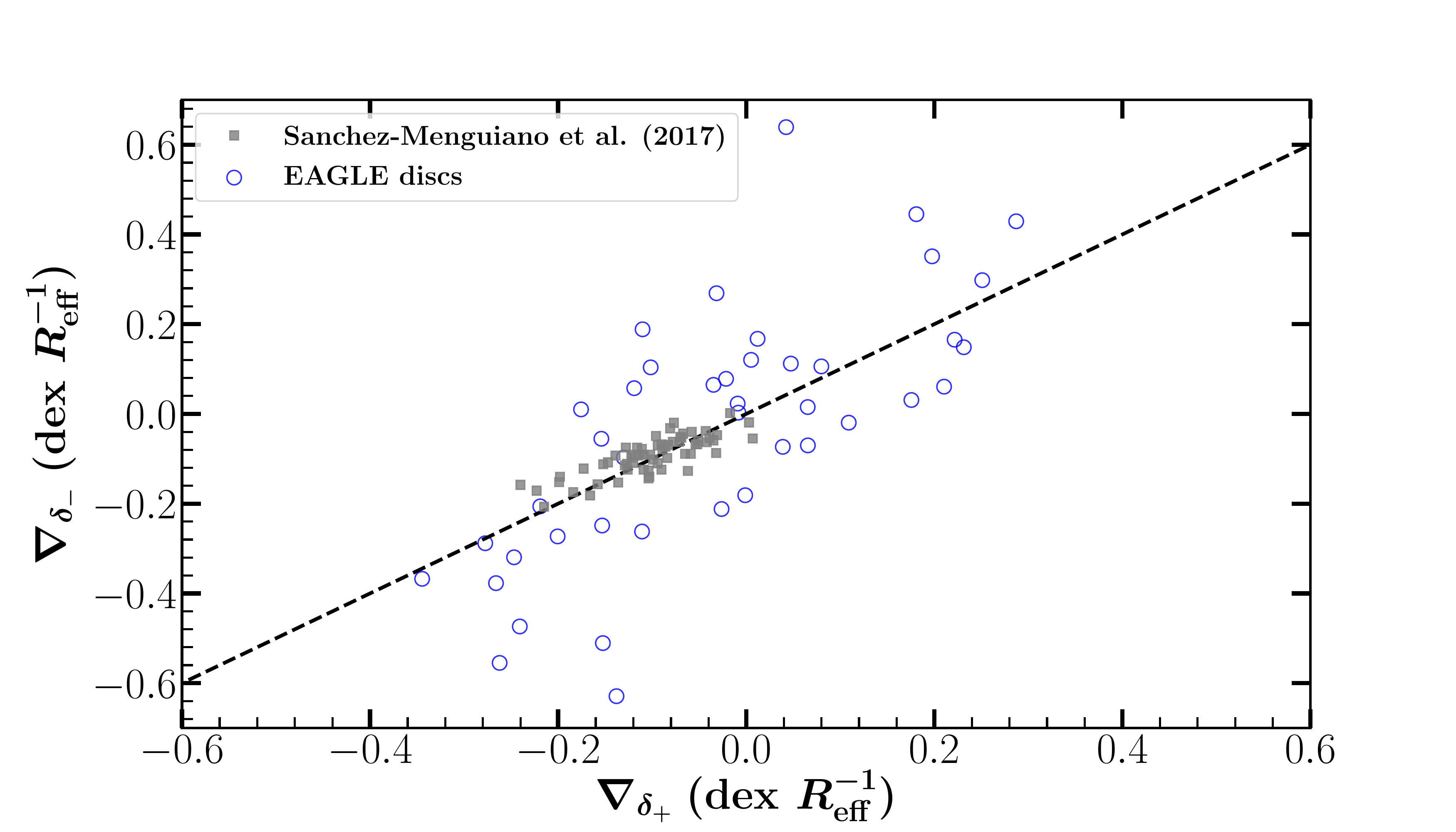}}
 \caption{$\nabla_{\rm \delta_{+}}$  as a function of $\nabla_{\rm \delta_{-}}$ defined by  super-young stars  (blue circles). The black dashed line corresponds to the 1:1 relation. Observations form \citet{SanchezMenguiano2017_1} have been included only for reference (gray points)  since they estimated the gradients by selected HII regions located in the arm and interarm regions.
}
\label{figsm}
\end{figure}

 Figure.~\ref{figsm} shows the relation of $\nabla_{\rm \delta_{-}}$ versus  $\nabla_{\rm \delta_{+}}$ together with the metallicity gradients reported by  \citet{SanchezMenguiano2017_1} are displayed for the sake of comparison. However, it is important to stress that  the latter were obtained by applying a different method. \citet{SanchezMenguiano2017_1} estimated
the metallicity gradients for HII regions located in the arm and interarm regions.
Nevertheless, both approaches intent to highlight the difference in the metallicity distribution of the SPs in the discs. It is clear that the simulated parameters show large dispersion which, at least in part, could be due to  inefficient mixing of the chemical elements or more clumpy star-formation activity. 

In order to complete the analysis, we explore if the smoothed abundances could reduce the scatter by repeating this analysis with the smoothed element abundances as shown in the Appendix. As can be seen from Fig.~\ref{5both}, using the smoothed abundances does not reduce the scatter in the metallicity gradients or  reduce the gap between the level of
 enrichment between the SPs classified as having metallicity excess or deficit.

\section{Conclusions}

We study the azimuthal variation of the oxygen abundances of young and super-young SPs which can be used as tracers of the
regions of active SF in the discs of galaxies selected from the higher resolution run of the EAGLE project, Ref-L025N0752.
The azimuthal variations of the metallicities can store information on the evolution of discs and currently, IFS observations provide detail information of the metallicity distributions. Additionally, they provide another route to confront the subgrid physics with observations. 
Our analysed sample is made of 106 (42) galaxies resolved with at least 1000 young (super-young) star particles. 
 As a consequence, our sample represents star-forming galaxies with SFR $>0.1~{\rm M\sun~yr^{-1}}$. 

Our main results can be summarized as follows:

\begin{itemize}

\item

Although we find a correlation between the metallicity gradients measured along a random direction and those estimated by using global
averages over the discs, the scatter is large enough to suggest azimuthal variations are significant.
The azimuthal dispersion of the slopes  and zero points of young stars in the EAGLE discs are found to be around  $0.12 \pm 0.03$ dex $R_{\rm eff}^{\rm -1}$ and    $0.12 \pm 0.03$ dex, respectively.

\item  A weak trend to have
larger metallicity azimuthal dispersions are detected for galaxies with lower star-formation activity and $D/T < 0.2$.
The larger variation for bulge-dominated galaxies and low star-formation activity is in agreement with the results obtained by \citet{Tissera2019_1}, where the gas-phase oxygen profiles
in the star formation regions were
analysed for galaxies of different morphologies in the EAGLE simulations.   Previous works reported bulge-dominated systems to have
larger frequency of mergers. These events could have destroyed part of the discs or disturbed them  by forming bar or spiral structures, which could have contributed to mix up the stellar populations.  It has also to be considered that these systems have lower star-formation activity and, even though we required a minimum number of stars to make the estimations, the dispersion could be more
affected by numerical resolution than those discs with larger star-formation rates. However, currently,  it is still difficult to have a large dataset of galaxies with a vast variety of morphologies and at the same time, with high numerical resolution.
Well-defined discs with $D/T >0.4$ show azimuthal dispersions of $~0.1$ dex $R_{\rm eff}^{\rm -1}$ in good agreement with the predictions by \citet{Spitoni2018_1}. We find no clear trends with $R_{\rm eff}$.

\item
The metallicity slopes estimated by using the super-young SPs with metallicity excess are slightly shallower than the overall estimations.
However, the  larger variation of these gradients in the simulations with respect to the current  observations \citep[e.g.][]{SanchezMenguiano2016_1} suggests the relevance of exploring in more detail the flow of metals at sub-galactic scales.

\end{itemize}


\section*{Acknowledgments}
{The authors thank to A. Benitez-Llambay and J.Schaye for useful comments.
JAHJ thanks to CONICYT, Programa de Astronom\'ia, Fondo ALMA-CONICYT 2017, C\'odigo de proyecto 31170038.
PBT acknowledges partial funding by Fondecyt Regular 2015 - 1150334 and Internal Project Unab 2019.
This project has received funding from the European Union Horizon 2020
Research and Innovation Programme under the Marie Sklodowska-Curie
grant agreement No 734374.
This worked used the RAGNAR cluster funded
by Fondecyt 1150334 and Universidad Andres Bello.
This work used the DiRAC Data Centric system at Durham University, operated by the Institute for Computational Cosmology on behalf of the STFC DiRAC HPC Facility (www.dirac.ac.uk). This equipment was funded by BIS National E-infrastructure capital grant ST/K00042X/1, STFC capital grants ST/H008519/1 and ST/K00087X/1, STFC DiRAC Operations grant ST/K003267/1 and Durham University. DiRAC is part of the National E-Infrastructure. We acknowledge PRACE for awarding us access to the Curie machine based in France at TGCC, CEA, Bruyeres-le-Chatel.}


\bibliographystyle{mnras}
\bibliography{bibliography}

\appendix

\section{Analysis of the metallicity dispersion}

\begin{figure}
\resizebox{8cm}{!}{\includegraphics{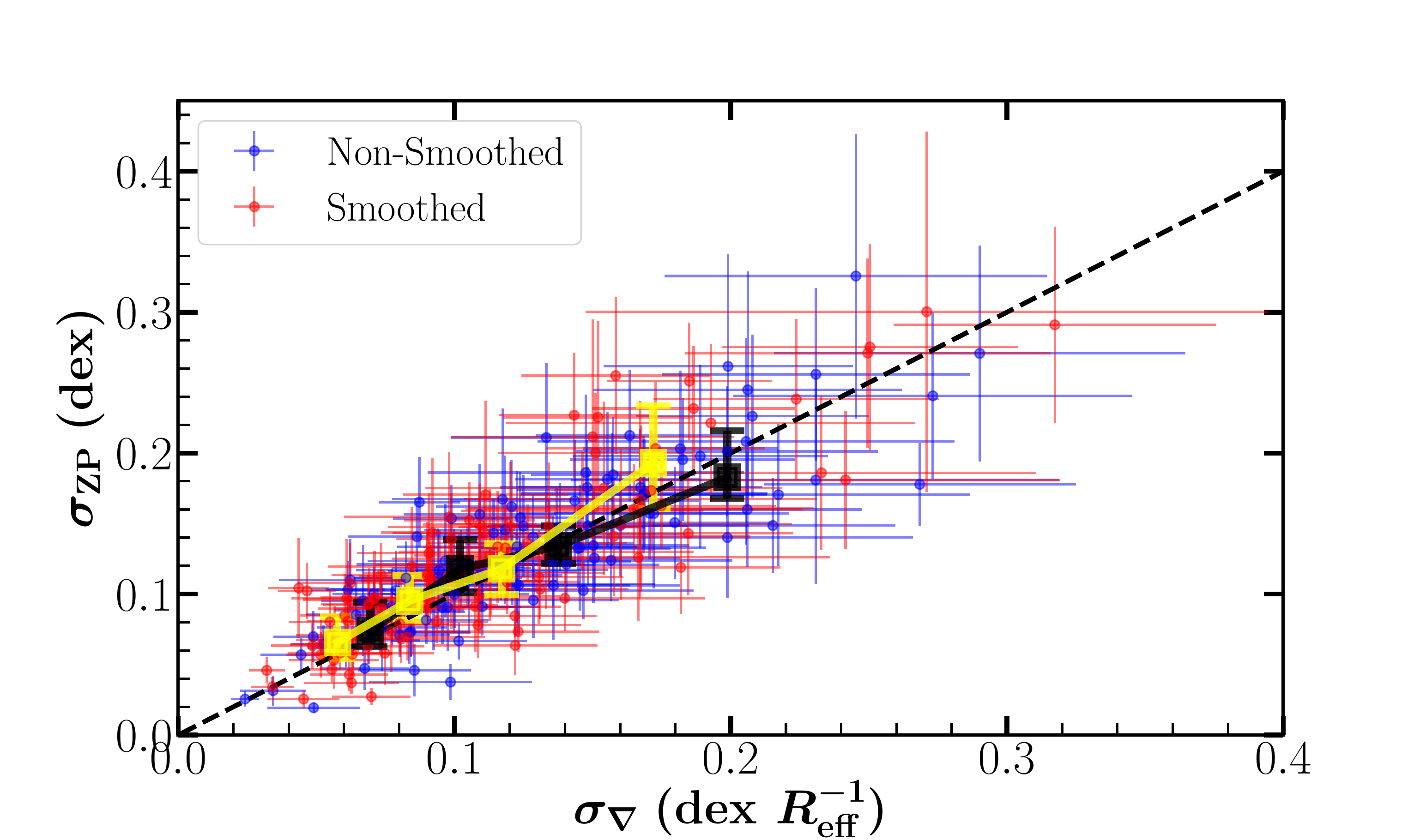}}
\caption{Median values of the standard deviations of the azimuthal zero points, $\sigma_{\rm ZP}$, as a function of those of the slopes, $\sigma_{\rm \nabla}$, for the non-smoothed (black squares)
and smoothed (yellow squares) metallicity profiles.The black (yellow) error bars
represent the 25 and 75 percentile of the non-smoothed (smoothed) relations. The blue (red) small dots and blue (red) error bars
represent the data for individual simulated galaxies and the corresponding errors obtained by applying a bootstrap sampling technique over the six azimuthal subregions.}
\label{3both}
\end{figure}

\begin{figure}
\resizebox{8cm}{!}{\includegraphics{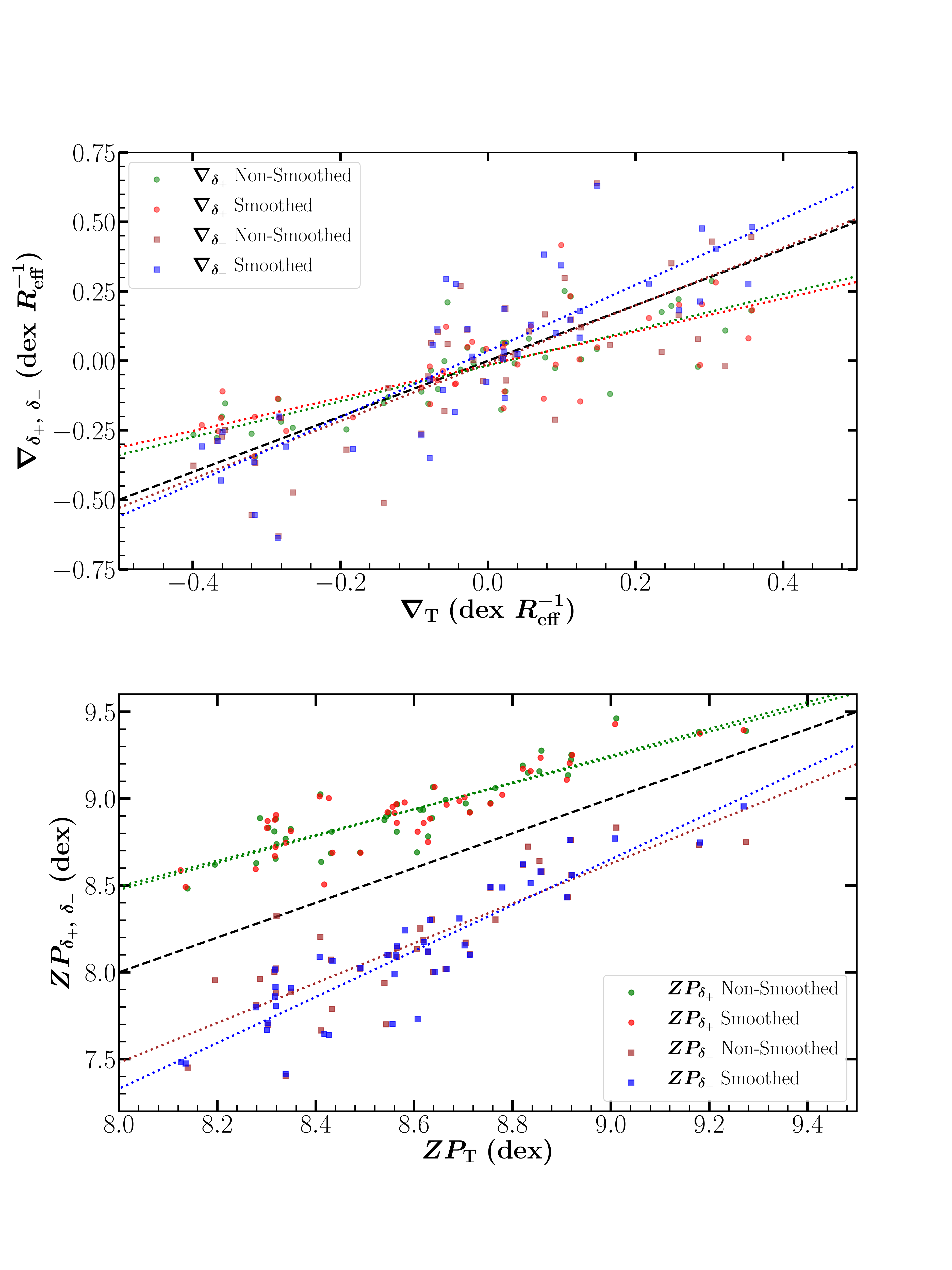}}
\caption{Upper panel:$\nabla_{\rm \delta_{+}}$ and $\nabla_{\rm \delta_{-}}$ as a function of $\nabla_{\rm T}$ estimated for the whole distribution of super-young stars using the non-smoothed and smoothed chemical abundances. The black dashed line represents the 1:1 relation. The dashed lines depict linear regressions to the  distributions of $\nabla_{\rm \delta_{+}}$  and $\nabla_{\rm \delta_{-}}$ for the smoothed and non-smoothed relations. Lower panel:$ZP_{\rm +}$ and  $ZP_{\rm -}$ as a function of  $ZP_{\rm T}$ . The black dashed lines represent a 1:1 relation, for comparison purposes.}
\label{5both}
\end{figure}

The EAGLE Project provides smoothed variables of the chemical abundances estimated by using the kernel function adopted for the Smoothed Particle Hydrodynamics  calculations.
These smoothed variables are estimated at the time the stars are formed. They  provide a rough estimation of the effects that a  more efficient mixing process might have.
We re-do all the calculations to assess if the trend changed when the smoothed abundances are used instead of the non-smoothed variables.   However, no significant or systematic differences are found. As an example, in Fig. ~\ref{3both} we show the median dispersion in the slope and zero point of the metallicity gradients for both cases.
Similarly,  in Fig.~\ref{5both} we show the slope and zero points obtained by using the star formation regions with excess or deficit metallicity with respect to the corresponding averages. As can be seen, there are no statistical differences between the results obtained from both relations.

\end{document}